\documentclass[prl,aps,twocolumn, showpacs, nofootinbib,superscriptaddress,notitlepage]{revtex4-1}
\usepackage{amssymb,amsthm,amsmath}
\usepackage{graphicx}   % figures
\usepackage{color}      % color is used in text
\usepackage{slashed}    % Feynman slash
\usepackage{epsfig}
\usepackage{subfigure}  % side-by-side figures
\usepackage{diagbox}
\usepackage{url}
\usepackage[normalem]{ulem}

\begin{document}

%\preprint{XXX}

\title{First Lattice QCD determination of  semileptonic decays of charmed-strange baryons $\Xi_c$}

\vspace{1.0cm}

\author{Qi-An Zhang}
\affiliation{ Key Laboratory for Particle Astrophysics and Cosmology (MOE), Shanghai Key Laboratory for Particle Physics and Cosmology, Tsung-Dao Lee Institute, Shanghai Jiao Tong University, Shanghai 200240, China}

\author{Jun Hua}
\affiliation{INPAC,  Key Laboratory for Particle Astrophysics and Cosmology (MOE),  Shanghai Key Laboratory for Particle Physics and Cosmology, School of Physics and Astronomy, Shanghai Jiao Tong University, Shanghai 200240, China}

\author{Fei Huang}
\affiliation{INPAC,  Key Laboratory for Particle Astrophysics and Cosmology (MOE),  Shanghai Key Laboratory for Particle Physics and Cosmology, School of Physics and Astronomy, Shanghai Jiao Tong University, Shanghai 200240, China}

\author{Renbo Li}
\affiliation{Nanjing Normal University, Nanjing, Jiangsu, 210023, China}

\author{Yuanyuan Li}
\affiliation{Nanjing Normal University, Nanjing, Jiangsu, 210023, China}

\author{Cai-Dian~L\"u } 
\affiliation{Institute of High Energy Physics, Chinese Academy of Sciences,
Beijing 100049, China}
\affiliation{ School of Physics, University of Chinese Academy of Sciences, Beijing 100049, China}

\author{Peng Sun}
\email{Corresponding author: sunpeng@njnu.edu.cn}
\affiliation{Nanjing Normal University, Nanjing, Jiangsu, 210023, China}

\author{Wei Sun}
\affiliation{Institute of High Energy Physics, Chinese Academy of Sciences,
Beijing 100049, China}

\author{Wei Wang}
\email{Corresponding author: wei.wang@sjtu.edu.cn}
\affiliation{INPAC,  Key Laboratory for Particle Astrophysics and Cosmology (MOE),  Shanghai Key Laboratory for Particle Physics and Cosmology, School of Physics and Astronomy, Shanghai Jiao Tong University, Shanghai 200240, China}

\author{Yi-Bo Yang}
\email{Corresponding author: ybyang@itp.ac.cn}
\affiliation{CAS Key Laboratory of Theoretical Physics, Institute of Theoretical Physics, Chinese Academy of Sciences, Beijing 100190, China}
\affiliation{School of Fundamental Physics and Mathematical Sciences, Hangzhou Institute for Advanced Study, UCAS, Hangzhou 310024, China}
\affiliation{International Centre for Theoretical Physics Asia-Pacific, Beijing/Hangzhou, China}

\date{\today}

\begin{abstract}
While the standard model is the most successfully theory to describe all interactions and constituents in elementary particle physics, it has been constantly examined for over  four decades. Weak decays of charm quarks can measure the coupling strength of quarks in different families and serve as an ideal probe for CP violation. As the   lowest charm-strange baryons with three different flavors,   $\Xi_c$ baryons (made of $csu$ or $csd$) have been extensively studied in experiments at the large hadron collider and in electron-positron collision. However the lack of reliable knowledge in theory becomes the unavoidable obstacle in the way. In this work, we use the state-of-the-art Lattice QCD techniques, and generate  2+1 clover fermion ensembles with two lattice spacings, $a=(0.108{\rm fm},0.080{\rm fm})$. 
 We then present the first {\it ab-initio}  lattice QCD determination of   form factors governing $\Xi_{c}\to \Xi \ell^+\nu_{\ell}$, analogous with the notable $\beta$-decay of nuclei. Our theoretical results for decay widths are consistent with and about two times more precise than the latest measurements by ALICE and Belle collaborations. Together with experimental measurements, we independently  determine the quark-mixing matrix element $|V_{cs}|$, which is found in good agreement with other determinations.  
\end{abstract}
\maketitle

{\it Introduction.}  
Since the establishment  in 1960s, the standard model   (SM) of particle physics has achieved many remarkable  successes, and has been constantly examined for over  four decades. Nowadays searching for new physics (NP) beyond the SM is the primary objective in particle physics, which usually  proceeds in two distinct directions. On the one side, new particles can be directly produced in high energy collisions for instance at the large hadron collider.  On the other side,  it is greatly valuable  to examine various low-energy  observables with prestigious  high precision that can give an indirect search for NP.  

Weak decays of heavy charm and bottom quarks provide an ideal platform to test the standard model  of particle physics, especially the Cabibbo-Kobayashi-Maskawa (CKM) paradigm which describes quark mixing and CP violation. Any significant deviation from   SM expectation for   CKM matrix  would indirectly   provide definitive clues for new physics beyond SM. Most of previous analysis concentrated on the meson sector like  $B$ and $D$ mesons, while recently heavy baryon decays started to determine $|V_{ub}/V_{cb}|$ from $\Lambda_b \to p \mu^-\bar{\nu}_\mu$ and $\Lambda_b \to \Lambda_c \mu^-\bar{\nu}_\mu$~\cite{Aaij:2015bfa}, and $|V_{cs}|$ from $\Lambda_c\to \Lambda e^+\nu_e$~\cite{Hinson:2004pj,Ablikim:2015prg}.

The study of weak decays of charmed baryons $\Xi_c^{+,0}$  especially $\Xi_c\to \Xi\ell^+\nu$ decays  is  greatly valuable  from various aspects.   First of all, the 
combination of  form factors from lattice QCD  (LQCD) and experimental results for branching fractions of semileptonic decays allows an independent determination of $|V_{cs}|$. Secondly,  a comparison of theory calculation  and   experimental measurements provides a stringent test of theoretical models. Thirdly, compared to the isosinglet counterpart $\Lambda_c$ whose decays have been extensively  in experiment~\cite{Ablikim:2015flg,Ablikim:2016tze,Ablikim:2016mcr,Ablikim:2017iqd,Ablikim:2018woi,Ablikim:2019zwe,Ablikim:2019hff,Zupanc:2013iki,Yang:2015ytm} and from LQCD~\cite{Meinel:2016dqj,Meinel:2017ggx}, the  iso-doublet $\Xi_c^{+,0}$  baryons  have more versatile decay modes. 
The closeness of decay branching fractions for  the exclusive $\Lambda_c\to \Lambda \ell^+\nu$ and   inclusive $\Lambda_c \to X\ell^+\nu$ modes~\cite{Ablikim:2018woi} shows a very different pattern with heavy bottom/charm mesons.  The study of  $\Xi_c\to \Xi \ell^+\nu$ decays and a combination  with $\Lambda_c$ decays can provide a way to validate this  pattern, which is valuable to understand the underlying dynamics in baryonic transition,  and test the flavor SU(3) symmetry~\cite{Lu:2016ogy,Grossman:2018ptn,Geng:2019bfz}. 
Moreover,   decays of $\Xi_c$ have played an important role in the study of doubly-charmed baryon $\Xi_{cc}^{++}$~\cite{Aaij:2018wzf}, precision measurement of the lifetime of $\Xi_b^0$~\cite{Aaij:2014esa}  and the discovery of  new exotic hadron candidates $\Omega_c$~\cite{Aaij:2017nav}.

Since the first observation of the inclusive semileptonic decay~\cite{Alexander:1994hp},   a number of  different  decay modes  of $\Xi_c$ have been studied in experiments~\cite{Aaij:2019kss,Aaij:2019lwg,Aaij:2020wtg,Aaij:2020wil,Li:2018qak}. In addition to measuring  branching fractions for suppressed modes,  the LHCb has also searched for CP violation in $\Xi_c^+\to   pK^-\pi^+$~\cite{Aaij:2020wil}.  Very recently, the ALICE~~\cite{ALICE} and Belle~\cite{2103.06496} collaborations  have measured the branching fractions for $\Xi_c\to \Xi\ell^+\nu$:
{\small
\begin{align}
 {\cal B}_{\rm ALICE}(\Xi_{c}^0\to \Xi^- e^+\nu_{e})=& (2.43\pm0.25\pm0.35\pm0.72) \%, \\
 {\cal B}_{\rm Belle}(\Xi_{c}^0\to \Xi^- e^+\nu_{e})
 =& (1.72\pm 0.10 \pm 0.12 \pm 0.50)\%  , \\
 {\cal B}_{\rm Belle}(\Xi_{c}^0\to \Xi^- \mu^+\nu_{\mu})
 =& (1.71\pm 0.17 \pm 0.13 \pm 0.50)\%,
 \end{align}}
where the last errors arise from the uncertainties in ${\cal B}(\Xi_{c}^0\to \Xi^-\pi^+)$~\cite{Li:2018qak}.

On theoretical side, the $\Xi_{c}\to \Xi$ transition  depends on six form factors which parametrize the matrix elements of vector and axial-vector currents between the $\Xi_{c}$ and $\Xi$ baryons.  Most available theoretical analyses of these form factors are based on   phenomenological  models~\cite{Zhao:2018zcb,Liu:2010bh,Azizi:2011mw,Geng:2018plk,Geng:2020gjh,Zhao:2021sje}, but  results  vary substantially depending on explicit assumptions. A first-principle calculation is extremely crucial for a precise determination of CKM matrix element, and  reliable analysis of CP violation in nonleptonic decays.  In this work,  we use the-state-of-the-art LQCD techniques and for the first time in the literature calculate  $\Xi_{c}\to \Xi$ form factors.  Predictions for semi-leptonic decay widths  are also presented, based on which the $|V_{cs}|$ is extracted.  As we will show, our results greatly improve the theoretical calculations, and are more precise than the experimental measurements. These results also serve as mandatory inputs for future  analysis of non-leptonic decays  particularly in the factorization scheme. 

{\it Lattice Setup.}   {This work is based on 2+1 flavor ensembles generated with tree level tadpole improved clover fermion action and tadpole improved Symanzik gauge action. One step of  Stout link smearing is applied to the gauge field used by the clover action to improve the stability of the pion mass for given bare quark mass. The tadpole improvement factors for quarks and gluons are tuned to the fourth root of the plaquette using Stout link smearing and the original gauge links. We start from the ensemble s108 with bare coupling $\beta=\frac{10}{g^2}=6.20$ and size $24^3\times 72$, determine the lattice spacing using Wilson flow \cite{Borsanyi:2012zs}, and tune the bare coupling for the s080 ensemble  with smaller lattice spacing to make their physical volume to be roughly the same. The information on the  two  ensembles used in this letter can be found in Tab.~\ref{table:s_set}}.

%%%%%%%%%%%%%%%%%%%%%
\begin{table}[htbp]
\begin{center}
\caption{Parameters of the 2+1 flavor clover fermion  ensembles used in this calculation. The $\pi$/$\eta_s$ masses and the lattice spacings are  given in units of MeV, and fm, respectively.}\label{table:s_set}
\begin{tabular}{ccccccccc}
\hline
  & $\beta=\frac{10}{g^2}$ & $L^3\times T$  &a & $c_{\textrm{sw}}$  & $\kappa_l$ & $m_{\pi} $ & $\kappa_s$&  $m_{\eta_s} $\\
\hline
s108&   6.20 & $24^3 \times  72$ & 0.108 &  1.161  & 0.1343 & 290 &  0.1330 & 640 \\
\hline
s080 &  6.41 & $32^3 \times  96$ & 0.080 &  1.141  & 0.1326 & 300 & 0.1318& 650 \\
\hline
 \hline
\end{tabular}
\end{center}
\end{table}
%%%%%%%%%%%%%%%%%%%%%

On these two ensembles, we use the charm quark mass $m_c^{s108}a=0.485$ and $m_c^{s080}a=0.235$, respectively, by requiring the corresponding  $J/\psi$ mass to have its physical value $m_{J/\psi}=3.96900(6)$GeV~\cite{Zyla:2020zbs} within 0.3\% accuracy. 

The extraction of $\Xi_c\to \Xi$  form factors requires the lattice QCD calculation of both three-point correlation function (3pt) from $\Xi_c$ to $\Xi$, and also the two point correlation functions (2pt) of both $\Xi_c$ and $\Xi$. The 3pt with weak current $J^\mu=V^\mu-A^\mu=\bar s \gamma^\mu(1-\gamma_5)c$ is defined by,
\begin{eqnarray}
C_{3}^{V-A}(q^2, t, t_{\rm seq}) &&= \int d^3\vec x d^3 \vec y e^{-i\vec p_{\Xi}\cdot \vec x} e^{-i\vec q\cdot \vec y}  T_{\gamma'\gamma}   \nonumber\\
&& \times \langle 0| \chi_{\gamma}^{\Xi}(\vec x, t_{\rm seq}) J^{\mu}(\vec y, t)\overline \chi_{\gamma'}^{\Xi_c}(\vec 0, 0)|0\rangle ,
\end{eqnarray}
where $\chi_{\gamma}^{\Xi, \Xi_c}$ is the interpolation field of $\Xi$ and $\Xi_c$, respectively, $T$ is a combination of the Dirac matrix that is chosen to project out the form factor. %\Red{YBY: Please provide the definition of T's we used in this work}
% $t_{\rm seq}$ denotes the source-sink separation in the time direction.
For the 2pt with $B=\Xi,\ \Xi_c$,
\begin{eqnarray}
C_{2}^{B}(t) = \int d^3\vec x e^{-i \vec p_{B} \cdot x} T^{\prime}_{\gamma'\gamma} \langle 0|\chi_{\gamma}^{B}(\vec x, t) \overline \chi_{\gamma'}^{B}(\vec 0,0) |0\rangle,
\end{eqnarray}
we choose $T^\prime$ as the identity matrix to simplify the expressions. Then we can define the following ratios for different projection matrices $T$ and current operator $V^{\mu}/A^{\mu}$,
{\small{
\begin{eqnarray}
R_{V/A}(T,\mu)= \sqrt{\frac{C_{3}^{V/A}(q^2, t, t_{\rm seq})C_{3}^{V/A}(q^2, t_{\rm seq}-t, t_{\rm seq})} {C_2^{B_1}(t_{\rm seq}) C_2^{B_2}(t_{\rm seq}) }  },  \label{eq:ratio}
\end{eqnarray}}}
where the subscript $V$ or $A$ corresponds to the vector or axial-vector current in the 3pt. After making use of the reduction formula, the ratios $R_F$ for the six form factors $F=(f_\perp, f_+, f_0, g_\perp, g_+, g_0)$ can be constructed by different combinations of $R_{V/A}(T, \mu)$. More details can be found in the supplemental material~\cite{supplemental}. 
Then we adopt the parameterization,
\begin{align}
R_F =& F\bigg(\frac{1+c_1e^{-\Delta E_1t}+c_2e^{-\Delta E_2 (t_{\rm seq}-t)}}{1+d_1e^{-\Delta E_1t_{\rm seq}}}\nonumber\\&\ \ \ \frac{1+c_1e^{-\Delta E_1 (t_{\rm seq}-t)}+c_2e^{-\Delta E_2t}}{1+d_2e^{-\Delta E_2t_{\rm seq}}}\bigg)^{1/2} \label{eq:R_parametrization}\\
\simeq& F[1+ c_1' (e^{-\Delta E_1t/2}+ e^{-\Delta E_1 (t_{\rm seq}-t)})/2], \label{eq:R_parametrization_b}
\end{align}
to eliminate excited-state contaminations and obtain the desired form factor $F$, where $\Delta E_1$ and $\Delta E_2$ ($>\Delta E_1$ ) describe the mass differences between the first excited states and ground states in the initial and final interpolation fields. It should be noted that  Eq.~(\ref{eq:R_parametrization})  is employed in the fit for most  cases, while Eq.~(\ref{eq:R_parametrization_b}) is used for large negative $q^2$ in combination with the ensemble s080 since the lattice results are noisy. We have checked that in these cases using Eq.~(\ref{eq:R_parametrization}) will lead to  consistent central values. 

%\Red{(We should justify this choose by applying Eq.~(\ref{eq:R_parametrization_b}) to the cases with better signal. If the result is consistent with that using Eq.~(\ref{eq:R_parametrization}) with comparable uncertainty, then the Eq.~(\ref{eq:R_parametrization_b}) is acceptable. If we are catching time, it would be left to the revised version.)}

\begin{figure}[!th]
\begin{center}
\includegraphics[width=0.45\textwidth]{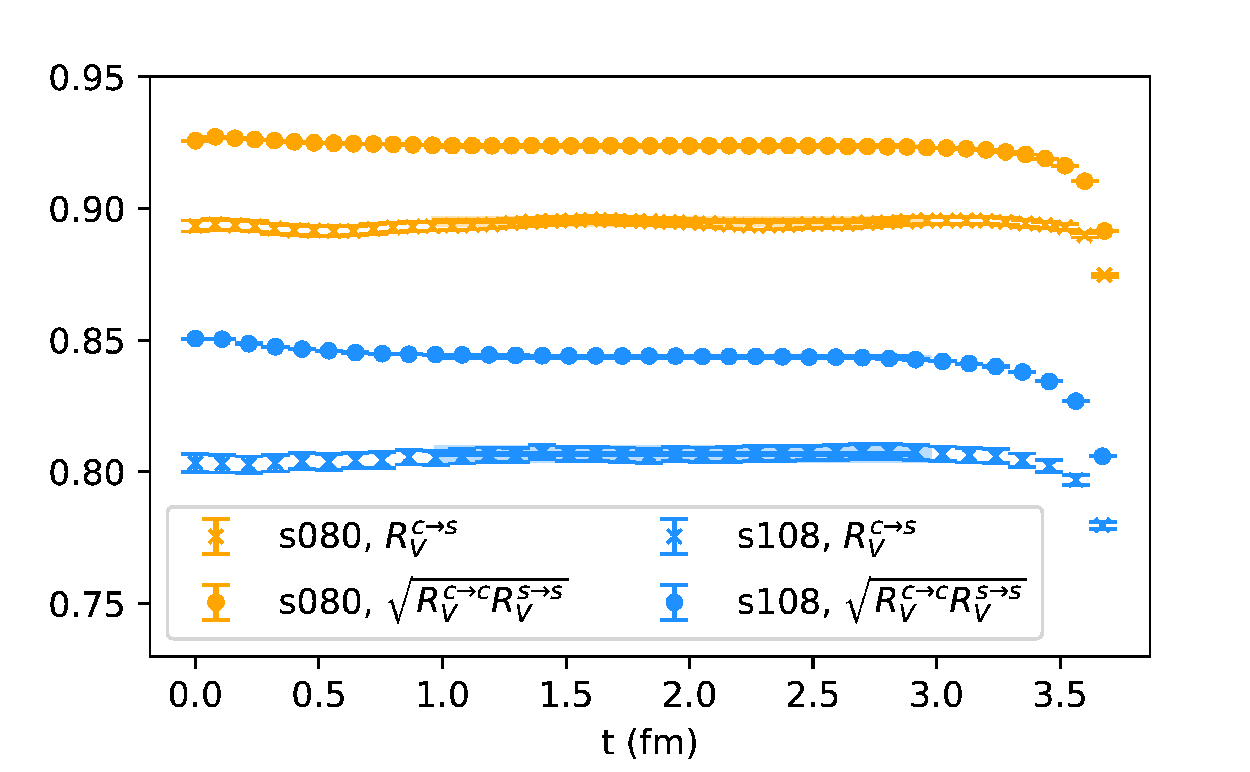}
\caption{Lattice results for $R_V^{c\to s}$ and $\sqrt{R_V^{c\to c}R_V^{s\to s}}$. The bands correspond to the ground state contributions $Z_V^{c\to s}$ and $\sqrt{Z_V^{c\to c}Z_V^{s\to s}}$ on the s080 and s108 ensembles, respectively. }\label{fig:npr}
\end{center}
\end{figure}

 The vector and axial-vector $c\to s$ currents on the lattice suffer from finite renormalization. Such a renormalization can be defined by the ratio of the conserved-vector-current-like $V_{\rm c}$ and the local current $V$ in the hadron matrix element,
\begin{align}
R^{q_1\to q_2}_V(t)&=\frac{\langle M_1(T/2)\sum_{\vec{x}}V^{q_1\to q_2}_{\rm c}(\vec{x},t)M_2(0)\rangle}{\langle M_1(T/2)\sum_{\vec{x}}V^{q_1\to q_2}(\vec{x},t)M_2(0)\rangle}\nonumber\\
&=Z^{q_1\to q_2}_V+{\cal O}(e^{-T/4 \Delta E}),
\end{align}
where $M_{1,2}$ are arbitrary pseudoscalar states and $\Delta E$ is the mass gap between the ground state and first exited state. For the  $c\to s$ current, one can use either the combination $(M_1,M_2)=(\eta_s, D_s)$, or the geometric average of those of the $s\to s$ current and $c \to c$ current using $(M_1,M_2)=(\eta_s, \eta_s)$ and $(\eta_c,\eta_c)$, respectively. We illustrate the $Z_V$ in Fig.~\ref{fig:npr}, in which the crosses and dots correspond to $R^{c\to s}_V(t)$ and $\sqrt{R^{c\to c}_V(t)R^{s\to s}_V(t)}$, respectively. Constant fits can describe the data at medium large $t\sim T/4$ well, and the difference between two definitions becomes smaller for the finer s080 ensemble  (upper yellow data), and both of them are also closer to one compared to the values for the coarser s108 ensemble  (lower blue data). Thus the differences between the two strategies arise from discretization effects. In the following discussion, we will use $Z^{c\to s}_V$ to obtain the central values of the final result, then repeat the analysis with $\sqrt{Z^{c\to c}_VZ^{s\to s}_V}$ and treat the differences as a systematic uncertainty.
%As in Ref.~\cite{xxx}, the needed vector normalization factor can be obtained the geometric average on those of the $s\to s$ current and $c \to c$ currents,
%\begin{align}
%Z_V^{c\to s}&=\sqrt{Z_V^{c}Z_V^{s}}\\
% Z_V^{q=c,s}&=\frac{2\langle O_{\eta_q}(T/2)\sum_{\vec{x}}\big(O_{\Gamma}(z;(\vec{x},T/4))\big)O_{\eta_q}^{\dagger}(0)\rangle}{\langle O_{\eta_q}(T/2)O_{\eta_q}^{\dagger}(0)\rangle},\nonumber
%\end{align}
%and we obtained  $Z_V$=0.8558(1) and 0.9305(1) on the s108 and s080 ensembles respectively.
Due to the chiral symmetry breaking of the clover fermion action, the renormalization factor of the axial-vector current is not exactly the same as for the vector one. Thus we use the off-shell quark matrix elements to define $Z_A$ as,
\begin{align}
Z^{c\to s}_A\equiv Z^{c\to s}_V\sqrt{ \frac{\mathrm{Tr}[\langle c|V^{\mu}|c\rangle\gamma^{\mu}\gamma_5]}{\mathrm{Tr}[\langle c|A^{\mu}|c\rangle\gamma^{\mu}]} \frac{\mathrm{Tr}[\langle s|V^{\mu}|s\rangle\gamma^{\mu}\gamma_5]}{\mathrm{Tr}[\langle s|A^{\mu}|s\rangle\gamma^{\mu}]}},
\end{align}
with multiple off-shell quark momenta $p^2$. With $a^2p^2$ extrapolation using three values of $p^2$ in the range of $a^2p^2\in[4,8]$, we obtained ${Z_A}/{Z_V}$= 1.010231(69) and 1.020296(68) on s108 and s080, respectively.

\begin{figure}[!th]
\begin{center}
\includegraphics[width=0.52\textwidth]{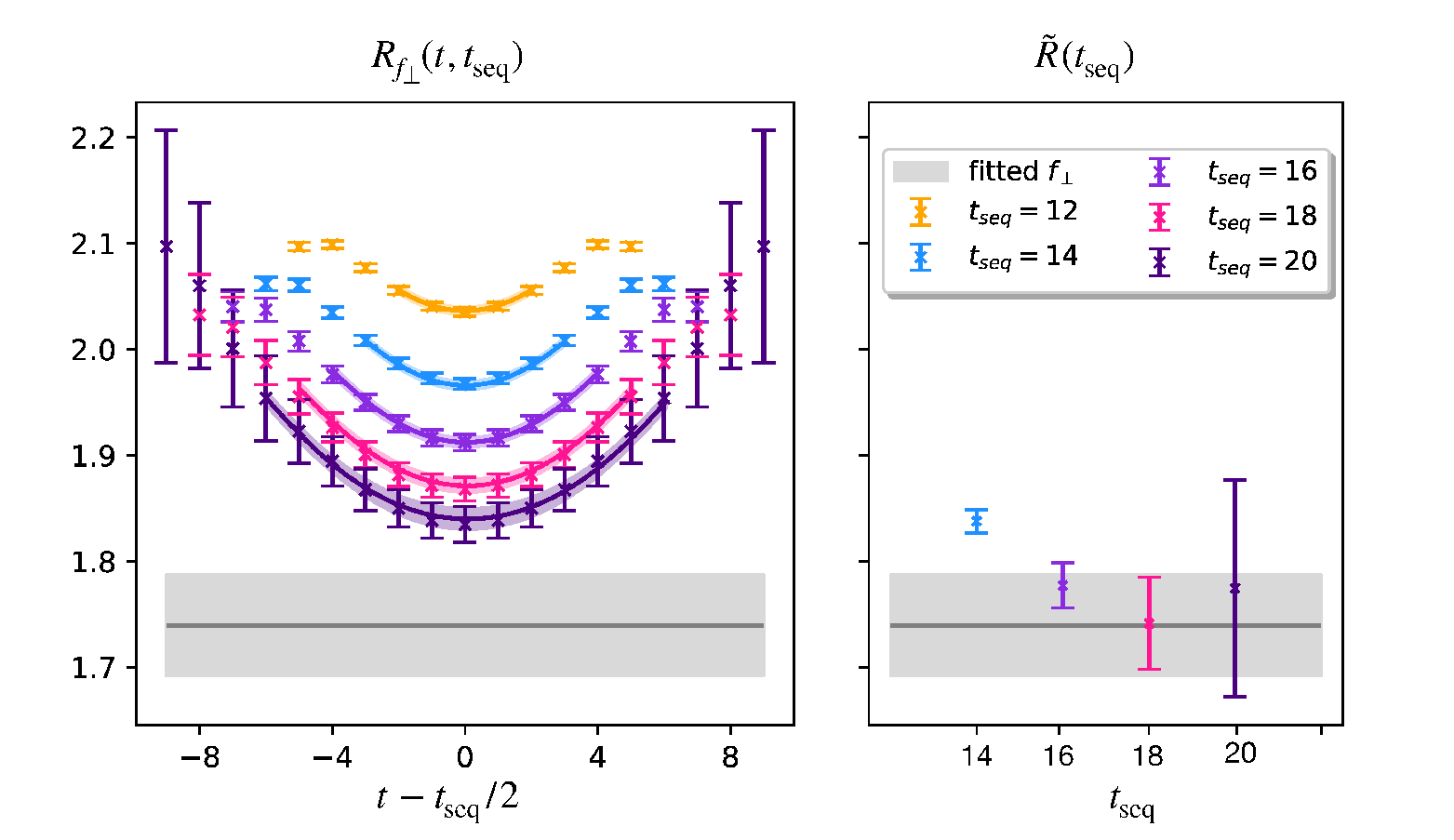}
\caption{Lattice results for the $f_\perp(\Xi_c\to \Xi)$ form factor on the s080 ensemble with $\vec{p}_{\Xi}=(0,0,1)\times \frac{2\pi}{La}$, in the source-sink separation range $[12a, 20a]$.  The left panel shows a two-state fit with the excited state contamination using the parametrization defined in Eq.~\eqref{eq:R_parametrization}, and the right panel gives the differential summed ratio. The ground-state matrix element (the grey band) obtained from the two-state fit agree with the differential summed ratio well when $t_{\rm seq}>14$.}\label{fig:fit}
\end{center}
\end{figure}

{\it Numerical Results.}
By choosing different reference time slices, we preform 48$\times$393 measurements on the s108 ensemble, and 72$\times$436 measurements on the s080 ensemble. The lattice results for the ratios $R_{f_\perp}$ with $\vec{p}_{\Xi}=(0,0,1)\times \frac{2\pi}{La}(\frac{2\pi}{La}\simeq 0.48{\rm GeV})$  are shown  in Fig.~\ref{fig:fit}. The $\chi^2/d.o.f$ are below/close to 1 for most fits of  400 bootstrap samples, which indicates a good fit quantity, and the colored bands in the left panel of Fig.~\ref{fig:fit} predicted by the fit agree with the data points well. To further validate the results, we  calculate  the differential summed ratio~\cite{Chang:2018uxx},
\begin{align}
 \tilde{R}(t_{\rm seq})&\equiv \frac{SR(t_{\rm seq})-SR(t_{\rm seq}-\Delta t)}{\Delta t},
\end{align}
and show the results in the right panel of Fig.~\ref{fig:fit}, where $SR(t_{\rm seq})\equiv\sum_{t_c<t<t_{\rm seq}-t_c}R_F(t, t_{\rm seq})$, $t_c=3a$ is the requirement used in the fits to suppress higher excited states contributions. One can see that  $\tilde{R}(t_{\rm seq})$ agrees with the grey band from the two-state fit well when $t_{\rm seq}>14a$.

%%%%%%%%%%%%%
%%%%%%%%%%%%%
%%%%%%%%%%%%%
%%%%%%%%%%%%%

%%%%%%%%%%%%%%%%%
\begin{figure*}
\begin{center}
\includegraphics[width=0.8\textwidth]{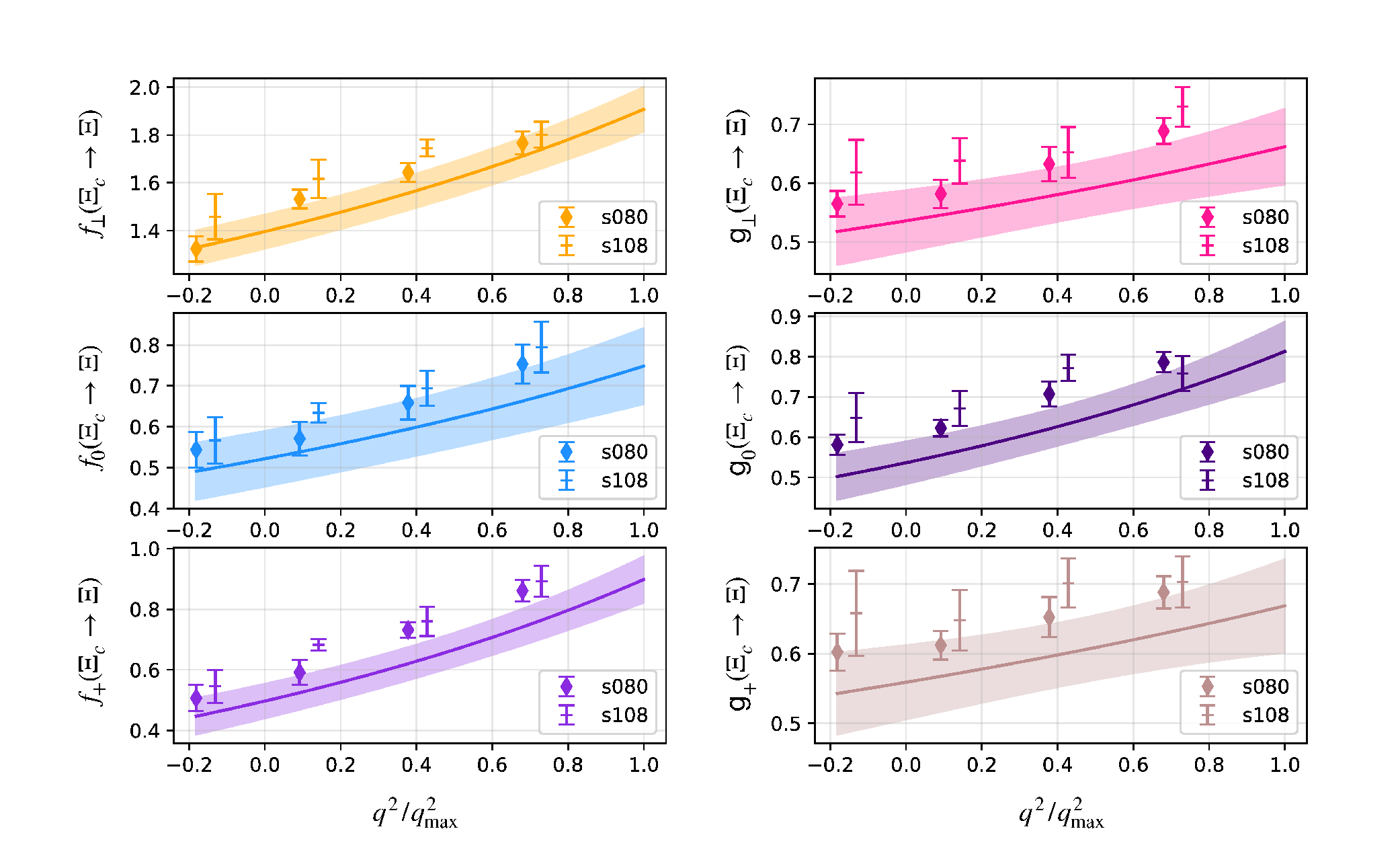}
\caption{The $q^2$ distribution for the $\Xi_c\to \Xi$ form factors. The $z$ expansion approach has been used to fit the lattice data. An extrapolation to the continuum limit has been made, and the shaded regions correspond to the final results with $a\to 0$.   }\label{fig:q2_formfactor}
\end{center}
\end{figure*}
%%%%%%%%%%%%%%%%%

%As  shown in Fig.~\ref{fig:q2_formfactor}, our results on the s108 ensemble are consistent with those on the s80 ensemble. 
To access the $q^2$ distribution, we employ the $z$-expansion parametrization of form factors that arises from analyticity and unitarity~\cite{Bourrely:2008za}
{\small
\begin{eqnarray}
f(q^2) = \frac{1}{1-q^2/(m_{\rm pole}^{f})^2} \sum_{n=0}^{n_{\rm max}} (c_{n}^f+ {d^f_{n}a^2})   [z(q^2)]^n,
\end{eqnarray}}
where
\begin{eqnarray}
z(q^2) = \frac{\sqrt{t_+-q^2} - \sqrt{t_+-t_0}}{\sqrt{t_+-q^2} + \sqrt{t_+-t_0}}. 
\end{eqnarray}
$t_0=q_{\rm max}^2= (m_{\Xi_c}-m_{\Xi})^2$ and $t_+ = (m_D+m_K)^2$, and $d^f_{n}$ describes the discretization error of each $z$-expansion parameter $c_{n}^f$.  The pole masses in the form factor are  $m_{\rm pole}^{f_+,f_\perp}=2.112$ GeV, $m_{\rm pole}^{f_0}=2.318$ GeV, $m_{\rm pole}^{g_+,g_\perp}=2.460$ GeV, and $m_{\rm pole}^{g_0}=1.968$ GeV.    We collect the fitted parameters in Tab.~\ref{table:form-factor}, and show the $q^2$ dependent form factor in the continuum limit (by eliminating the $d^f_{n}$ terms) from the fit and also the Lattice results at given $q^2$ in Fig.~\ref{fig:q2_formfactor}. As shown in the figure, our form factor results for the s108 ensemble and the s080 ensemble show only small discretization effects.

%%%%%%%%%%%%%%%%%%%%%
\begin{table}[htbp]
\begin{center}
\caption{\label{table:form-factor}
Results for the $z$-expansion parameters describing the form factors with statistical errors. }
\begin{tabular}{cccccc}
\hline
   &  &  $c_0$ & $c_1$ & $c_2$ & \\
 \hline
 & $f_{\perp}$    & 1.51$\pm$0.09 & -1.88$\pm$1.21 & 1.71$\pm$0.49 & \\
 & $f_{0}$         & 0.64$\pm$0.09 & -1.83$\pm$1.22 & 0.56$\pm$0.51 & \\
 & $f_+$           & 0.77$\pm$0.07 & -4.09$\pm$1.18 & 0.35$\pm$0.49 & \\
 & $g_{\perp}$   & 0.56$\pm$0.07 & -0.35$\pm$1.26 & 0.15$\pm$0.29 & \\
 & $g_{0}$       & 0.63$\pm$0.07 & -1.37$\pm$1.36 & 0.15$\pm$0.29 & \\
 & $g_+$          & 0.56$\pm$0.08 & 0.00$\pm$1.38 & 0.14$\pm$0.29 & \\
 \hline
 \hline
\end{tabular}
\end{center}
\end{table}
%%%%%%%%%%%%%%%%%%%%%

%%%%%%%%%%%%%%
\begin{figure}[!th]
\begin{center}
\includegraphics[width=0.45\textwidth]{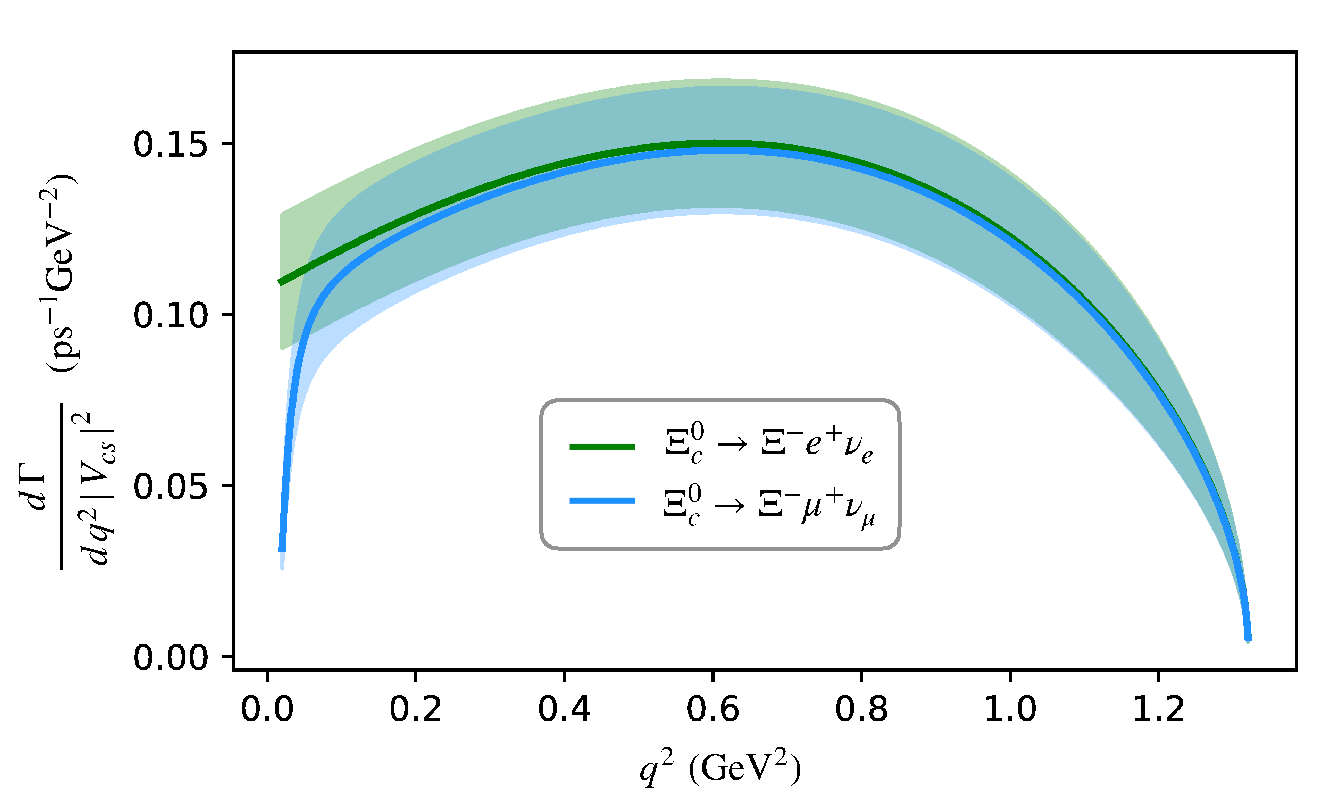}
\caption{Predictions for the differential decay widths for the $\Xi_{c}^0\to \Xi^- e^+\nu_e$ and $\Xi_{c}^0\to \Xi^- \mu^+\nu_\mu$, divided by $|V_{cs}|^2$ in units of ${\rm ps}^{-1} {\rm GeV}^{-2}$. }\label{fig:decay_width}
\end{center}
\end{figure}
%%%%%%%%%%%%%%

In Fig.~\ref{fig:decay_width}, we use the above form factors to predict differential decay widths (in units of ${\rm ps}^{-1} {\rm GeV}^{-2}$) for $\Xi_{c}^0\to \Xi^- \ell^+\nu$ divided by $|V_{cs}|^2$ as a function of $q^2$.  Results for $\Xi_{c}^+\to \Xi^0 \ell^+\nu$ are also similar. 
Using the lifetime from PDG: $\tau({\Xi^0_c})= (1.53\pm0.06)\times 10^{-13}$s and  $\tau({\Xi^-_c})= (4.56\pm0.05)\times 10^{-13}$s, and $|V_{cs}|=0.97320\pm0.00011$~\cite{Zyla:2020zbs}, one can obtain the decay branching fractions:
\begin{align}
{\cal B}(\Xi_{c}^0\to \Xi^- e^+\nu_e)=&2.38(0.30)_{\mathrm{stat.}}(0.32)_{\mathrm{ext.}}(0.07)_{\mathrm{ren.}}\% ,\nonumber\\
{\cal B}(\Xi_{c}^0\to \Xi^- \mu^+\nu_\mu)=& 2.29(0.29)_{\mathrm{stat.}}(0.30)_{\mathrm{ext.}}(0.06)_{\mathrm{ren.}}\%,\nonumber\\
{\cal B}(\Xi_{c}^+\to \Xi^0 e^+\nu_e)=&7.18(0.90)_{\mathrm{stat.}}(0.96)_{\mathrm{ext.}}(0.20)_{\mathrm{ren.}}\% ,\nonumber\\
{\cal B}(\Xi_{c}^+\to \Xi^0 \mu^+\nu_\mu)=&6.91(0.87)_{\mathrm{stat.}}(0.91)_{\mathrm{ext.}}(0.19)_{\mathrm{ren.}}\% .
\end{align}
The first errors come  from statistical fluctuations, while the second and third ones are systematic uncertainties arised from the differences between continuum-extrapolated results  and the ones using the s080 ensemble, and  the differences between the results using $Z_{V/A}^{c\to s}$ or $\sqrt{Z_{V/A}^{c\to c}Z_{V/A}^{s\to s}}$ in the renormalization, respectively.   Our predictions for branching fractions are consistent with model predictions in  Ref.~\cite{Zhao:2018zcb}, but  smaller than the ones in Ref.~\cite{Liu:2010bh,Geng:2020gjh}.  Compared to  the previous theoretical  results which  have  typically  $30\%\sim 50\%$ parametric uncertainties and uncontrollable systematic uncertainties,  our results have greatly improved the theoretical predictions.  Our calculation also indicates   sizable SU(3) symmetry breaking effects compared to $\Lambda_c\to \Lambda\ell^+\nu$ decays~\cite{Ablikim:2015prg,Meinel:2016dqj}.

The ratio of branching fractions is predicted as:
\begin{eqnarray}
R_{\mu/e} &&= \frac{{\cal B}(\Xi_{c}^0 \to \Xi^- \mu^+\nu_\mu)}{{\cal B}(\Xi_{c}^0\to \Xi^- e^+\nu_e)} = \frac{{\cal B}(\Xi_{c}^+ \to \Xi^0 \mu^+\nu_\mu)}{{\cal B}(\Xi_{c}^+\to \Xi^0 e^+\nu_e)} \nonumber\\
&&= 0.962(0.003)_{\mathrm{stat.}}(0.002)_{\mathrm{syst.}}, \label{eq:ratio_pre}
\end{eqnarray}
where most uncertainties from form factors have cancelled to a large extent. The deviation from unity arises from the mass differences between  muon and   electron.  This result  is   consistent with and much more precise than the Belle measurement: $R_{\mu/e} =1.00\pm0.11\pm0.09$~\cite{2103.06496}, which indicates that effects not covered by our lattice calculation are probably less significant at our level of precision.

{Our results for branching fractions are consistent with and about two times more precise than the measurements   by ALICE and Belle collaborations as shown in Eq.~(1-3).  Using the ALICE measurement~\cite{ALICE}, we give a determination of $|V_{cs}|$:
 \begin{eqnarray}
|V_{cs}|=0.983(0.060)_{\mathrm{stat.}}(0.065)_{\mathrm{syst.}}(0.167)_{\mathrm{exp.}},
 \end{eqnarray}
 where the first two uncertainties are statistical and systematic uncertainties of the theoretical results, and the last ones are dominant and come from experimental data.
Using the Belle result~\cite{2103.06496}, we also have: 
\begin{eqnarray}
|V_{cs}|= 0.834(0.051)_{\mathrm{stat.}}(0.056)_{\mathrm{syst.}}(0.127)_{\mathrm{exp.}},
\end{eqnarray}
which is obtained by combing $\Xi_{c}^0\to \Xi^- e^+\nu_e$ and $\Xi_{c}^0\to \Xi^- \mu^+\nu_\mu$. Using the individual channel, we have $|V_{cs}|_{(\ell=e)}=0.830(0.051)_{\mathrm{stat.}}(0.055)_{\mathrm{syst.}}(0.128)_{\mathrm{exp.}}$ and  $|V_{cs}|_{(\ell=\mu)}=0.846(0.052)_{\mathrm{stat.}}(0.056)_{\mathrm{syst.}}(0.135)_{\mathrm{exp.}}$. 
Both results of $|V_{cs}|$ from ALICE and Belle data are consistent with the global fit~\cite{Zyla:2020zbs}  within 1-$\sigma$. }

It is necessary to point out that  the largest errors in the extracted results for $|V_{cs}|$ are from   experimental data on  ${\cal B}(\Xi_c^0\to \Xi^-\pi^+)$~\cite{Li:2018qak}. This can be improved by more precise measurements at   LHCb, Belle-II, BESIII and other experiments in future.   It should also be noted that as a conservative estimate, we have included systematic uncertainties (about $6\%$). In the  continuum extrapolation, the statistical uncertainties in the two lattice ensembles are added  and the final  uncertainties are also about $6\%$.

{\it Conclusions.}  We have reported the first lattice QCD calculation of the form factors governing the $\Xi_{c}\to \Xi \ell^+\nu_{\ell}$ at two lattice spacings and extrapolated them to the continuum. 
Using the CKM matrix element $|V_{cs}|$ from PDG and the $\Xi_{c}$ lifetimes, we predict the branching fractions  ${\cal B}(\Xi_{c}^0\to \Xi^- e^+ \nu_{e})=2.38(0.30)_{\mathrm{stat.}}(0.32)_{\mathrm{syst.}}\% $, ${\cal B}(\Xi_{c}^0\to \Xi^- \mu^+ \nu_{\mu})=2.29(0.29)_{\mathrm{stat.}}(0.31)_{\mathrm{syst.}}\% $, ${\cal B}(\Xi_{c}^+\to \Xi^0 e^+ \nu_{e})= 7.18(0.90)_{\mathrm{stat.}}(0.98)_{\mathrm{syst.}}\%$, and ${\cal B}(\Xi_{c}^+\to \Xi^0 \mu^+ \nu_{\mu})= 6.91(0.87)_{\mathrm{stat.}}(0.93)_{\mathrm{syst.}}\%$ with both statistical and systematic uncertainties. 
Our results have greatly improved previous  theoretical calculations, and are consistent with and about two times more precise than the measurements   by ALICE and Belle collaborations. Our calculation also indicates   sizable SU(3) symmetry breaking effects compared to $\Lambda_c\to \Lambda\ell^+\nu$ decays. 
These results also serve as mandatory inputs for the analysis of non-leptonic decays in the factorization scheme. Using the measured branching fraction from two experiments together with our lattice results, we determine the CKM matrix element $|V_{cs}|=0.983(0.060)_{\mathrm{stat.}}(0.065)_{\mathrm{syst.}}(0.167)_{\mathrm{exp.}}$ and $0.834(0.051)_{\mathrm{stat.}}(0.056)_{\mathrm{syst.}}(0.127)_{\mathrm{exp.}}$, where the errors come from the theoretical and experimental uncertainties, respectively.

\vspace{0.5cm}

{\it Acknowledgment.}---
We thank Andreas Sch\"afer  for valuable discussions,   Y.B. Yin,  J.~Zhu and T.~Cheng for pointing out the ALICE result in Ref.~\cite{ALICE}, C.P. Shen and Y.B. Li for the correspondence on the Belle measurement~\cite{2103.06496}, and  W.B.~Qian, Y.H.~Xie, H.B.~Li, B.Q.~Ke and X.R.~Lyu for noticing us the  experimental studies  of $\Xi_c$ decays at LHCb and BESIII. 
We greatly thank Prof. En-Ke Wang, Nu Xu and Rong-Gen Cai for their support when
the gauge configurations are generated on the cluster supported by Southern Nuclear Science Computing Center (SNSC) and also HPC Cluster of ITP-CAS.
The LQCD calculations were performed using the Chroma software suite~\cite{Edwards:2004sx} and QUDA~\cite{Clark:2009wm,Babich:2011np,Clark:2016rdz} through HIP programming model~\cite{Bi:2020wpt}.
The numerical calculation is supported by Chinese Academy of Science CAS Strategic Priority Research Program of Chinese Academy of Sciences, Grant No. XDC01040100.
The setup for numerical simulations was conducted  on the $\pi$ 2.0 cluster supported by the Center for High Performance Computing at Shanghai Jiao Tong University.
This work is supported in part by Natural Science Foundation of China under grant Nos. 11735010,  U2032102, 11653003,  12005130, 11521505, 12070131001,  11975127 and 11935017,  Natural Science Foundation of Shanghai under grant No. 15DZ2272100, the China Postdoctoral Science Foundation and the National Postdoctoral Program for Innovative Talents (Grant No. BX20190207), National Key Research and Development Program of China under Contract No. 2020YFA0406400,  Jiangsu Specially Appointed Professor Program, the Strategic Priority Research Program of Chinese Academy of Sciences, Grant No. XDB34030300, and  a NSFC-DFG joint grant under grant No. 12061131006 and SCHA 458/22.


\begin{thebibliography}{99} 

 


%\cite{Aaij:2015bfa}
\bibitem{Aaij:2015bfa}
R.~Aaij \textit{et al.} [LHCb],
%``Determination of the quark coupling strength $|V_{ub}|$ using baryonic decays,''
Nature Phys. \textbf{11}, 743-747 (2015),
doi:10.1038/nphys3415
[arXiv:1504.01568 [hep-ex]].
%155 citations counted in INSPIRE as of 10 Feb 2021

%\cite{Hinson:2004pj}
\bibitem{Hinson:2004pj}
J.~W.~Hinson \textit{et al.} [CLEO],
%``Improved measurement of the form-factors in the decay Lambda+(c) ---\ensuremath{>} Lambda e+ nu(e),''
Phys. Rev. Lett. \textbf{94}, 191801 (2005),
doi:10.1103/PhysRevLett.94.191801
[arXiv:hep-ex/0501002 [hep-ex]].
%23 citations counted in INSPIRE as of 22 Jan 2021

%\cite{Ablikim:2015prg}
\bibitem{Ablikim:2015prg}
M.~Ablikim \textit{et al.} [BESIII],
%``Measurement of the absolute branching fraction for $\Lambda^+_{c}\to \Lambda e^+\nu_e$,''
Phys. Rev. Lett. \textbf{115},  221805 (2015),
doi:10.1103/PhysRevLett.115.221805
[arXiv:1510.02610 [hep-ex]].
%59 citations counted in INSPIRE as of 17 Feb 2021

%\cite{Ablikim:2015flg}
\bibitem{Ablikim:2015flg}
M.~Ablikim \textit{et al.} [BESIII],
%``Measurements of absolute hadronic branching fractions of $\Lambda_{c}^{+}$ baryon,''
Phys. Rev. Lett. \textbf{116},  052001 (2016),
doi:10.1103/PhysRevLett.116.052001
[arXiv:1511.08380 [hep-ex]].
%122 citations counted in INSPIRE as of 11 Feb 2021

%\cite{Ablikim:2016tze}
\bibitem{Ablikim:2016tze}
M.~Ablikim \textit{et al.} [BESIII],
%``Measurement of Singly Cabibbo Suppressed Decays $\Lambda_c^{+}\to p\pi^{+}\pi^{-}$ and $\Lambda_c^{+}\to pK^{+}K^{-}$,''
Phys. Rev. Lett. \textbf{117},   232002 (2016),
doi:10.1103/PhysRevLett.117.232002
[arXiv:1608.00407 [hep-ex]].
%29 citations counted in INSPIRE as of 29 Jan 2021

%\cite{Ablikim:2016mcr}
\bibitem{Ablikim:2016mcr}
M.~Ablikim \textit{et al.} [BESIII],
%``Observation of $\Lambda^+_{c}\to nK^0_S\pi^+$,''
Phys. Rev. Lett. \textbf{118},   112001 (2017),
doi:10.1103/PhysRevLett.118.112001
[arXiv:1611.02797 [hep-ex]].
%29 citations counted in INSPIRE as of 22 Jan 2021

%\cite{Ablikim:2017iqd}
\bibitem{Ablikim:2017iqd}
M.~Ablikim \textit{et al.} [BESIII],
%``Observation of the decay $\Lambda_c^+\rightarrow \Sigma^- \pi^+\pi^+\pi^0$,''
Phys. Lett. B \textbf{772}, 388-393 (2017),
doi:10.1016/j.physletb.2017.06.065
[arXiv:1705.11109 [hep-ex]].
%20 citations counted in INSPIRE as of 22 Jan 2021
%\cite{Ablikim:2018woi}
\bibitem{Ablikim:2018woi}
M.~Ablikim \textit{et al.} [BESIII],
%``Measurement of the absolute branching fraction of the inclusive semileptonic $\Lambda_c^+$ decay,''
Phys. Rev. Lett. \textbf{121}, no.25, 251801 (2018)
doi:10.1103/PhysRevLett.121.251801
[arXiv:1805.09060 [hep-ex]].
%10 citations counted in INSPIRE as of 23 Mar 2021

%\cite{Ablikim:2019zwe}
\bibitem{Ablikim:2019zwe}
M.~Ablikim \textit{et al.} [BESIII],
%``Measurements of Weak Decay Asymmetries of $\Lambda_c^+\to pK_S^0$, $\Lambda\pi^+$, $\Sigma^+\pi^0$, and $\Sigma^0\pi^+$,''
Phys. Rev. D \textbf{100}, no.7, 072004 (2019),
doi:10.1103/PhysRevD.100.072004
[arXiv:1905.04707 [hep-ex]].
%11 citations counted in INSPIRE as of 22 Jan 2021

%\cite{Ablikim:2019hff}
\bibitem{Ablikim:2019hff}
M.~Ablikim \textit{et al.} [BESIII],
%``Future Physics Programme of BESIII,''
Chin. Phys. C \textbf{44},   040001 (2020),
doi:10.1088/1674-1137/44/4/040001
[arXiv:1912.05983 [hep-ex]].
%90 citations counted in INSPIRE as of 25 Feb 2021



%\cite{Zupanc:2013iki}
\bibitem{Zupanc:2013iki}
A.~Zupanc \textit{et al.} [Belle],
%``Measurement of the Branching Fraction $\mathcal B(\Lambda_c^+ \to p K^- \pi^+)$,''
Phys. Rev. Lett. \textbf{113},  042002 (2014),
doi:10.1103/PhysRevLett.113.042002
[arXiv:1312.7826 [hep-ex]].
%104 citations counted in INSPIRE as of 19 Feb 2021

%\cite{Yang:2015ytm}
\bibitem{Yang:2015ytm}
S.~B.~Yang \textit{et al.} [Belle],
%``First Observation of Doubly Cabibbo-Suppressed Decay of a Charmed Baryon: $\Lambda^{+}_{c} \rightarrow p K^{+} \pi^{-}$,''
Phys. Rev. Lett. \textbf{117},  011801 (2016), 
doi:10.1103/PhysRevLett.117.011801
[arXiv:1512.07366 [hep-ex]].
%40 citations counted in INSPIRE as of 25 Feb 2021

%\cite{Meinel:2016dqj}
\bibitem{Meinel:2016dqj}
S.~Meinel,
%``$\Lambda_c \to \Lambda l^+ \nu_l$ form factors and decay rates from lattice QCD with physical quark masses,''
Phys. Rev. Lett. \textbf{118},   082001 (2017), 
doi:10.1103/PhysRevLett.118.082001
[arXiv:1611.09696 [hep-lat]].
%26 citations counted in INSPIRE as of 22 Jan 2021

%\cite{Meinel:2017ggx}
\bibitem{Meinel:2017ggx}
S.~Meinel,
%``$\Lambda_c \to N$ form factors from lattice QCD and phenomenology of $\Lambda_c \to n \ell^+ \nu_\ell$ and $\Lambda_c \to p \mu^+ \mu^-$ decays,''
Phys. Rev. D \textbf{97},  034511 (2018), 
doi:10.1103/PhysRevD.97.034511
[arXiv:1712.05783 [hep-lat]].
%27 citations counted in INSPIRE as of 22 Jan 2021

%\cite{Lu:2016ogy}
\bibitem{Lu:2016ogy}
C.~D.~L\"u, W.~Wang and F.~S.~Yu,
%``Test flavor SU(3) symmetry in exclusive $\Lambda_c$ decays,''
Phys. Rev. D \textbf{93}, no.5, 056008 (2016)
doi:10.1103/PhysRevD.93.056008
[arXiv:1601.04241 [hep-ph]].
%75 citations counted in INSPIRE as of 23 Mar 2021
%\cite{Grossman:2018ptn}
\bibitem{Grossman:2018ptn}
Y.~Grossman and S.~Schacht,
%``U-Spin Sum Rules for CP Asymmetries of Three-Body Charmed Baryon Decays,''
Phys. Rev. D \textbf{99}, no.3, 033005 (2019)
doi:10.1103/PhysRevD.99.033005
[arXiv:1811.11188 [hep-ph]].
%18 citations counted in INSPIRE as of 23 Mar 2021

%\cite{Geng:2019bfz}
\bibitem{Geng:2019bfz}
C.~Q.~Geng, C.~W.~Liu, T.~H.~Tsai and S.~W.~Yeh,
%``Semileptonic decays of anti-triplet charmed baryons,''
Phys. Lett. B \textbf{792}, 214-218 (2019)
doi:10.1016/j.physletb.2019.03.056
[arXiv:1901.05610 [hep-ph]].
%17 citations counted in INSPIRE as of 23 Mar 2021


%\cite{Aaij:2018wzf}
\bibitem{Aaij:2018wzf}
R.~Aaij \textit{et al.} [LHCb],
%``Measurement of the Lifetime of the Doubly Charmed Baryon $\Xi_{cc}^{++}$,''
Phys. Rev. Lett. \textbf{121}, no.5, 052002 (2018)
doi:10.1103/PhysRevLett.121.052002
[arXiv:1806.02744 [hep-ex]].
%73 citations counted in INSPIRE as of 23 Mar 2021

%\cite{Aaij:2014esa}
\bibitem{Aaij:2014esa}
R.~Aaij \textit{et al.} [LHCb],
%``Precision measurement of the mass and lifetime of the $\Xi_b^0$ baryon,''
Phys. Rev. Lett. \textbf{113}, 032001 (2014)
doi:10.1103/PhysRevLett.113.032001
[arXiv:1405.7223 [hep-ex]].

%\cite{Aaij:2017nav}
\bibitem{Aaij:2017nav}
R.~Aaij \textit{et al.} [LHCb],
%``Observation of five new narrow $\Omega_c^0$ states decaying to $\Xi_c^+ K^-$,''
Phys. Rev. Lett. \textbf{118}, no.18, 182001 (2017)
doi:10.1103/PhysRevLett.118.182001
[arXiv:1703.04639 [hep-ex]].
%218 citations counted in INSPIRE as of 23 Mar 2021

%\cite{Alexander:1994hp}
\bibitem{Alexander:1994hp}
J.~P.~Alexander \textit{et al.} [CLEO],
%``First observation of Xi(c)+ ---\ensuremath{>} Xi0 e+ electron-neutrino and a measurement of the Xi(c)+ / Xi(c)0 lifetime ratio,''
Phys. Rev. Lett. \textbf{74}, 3113-3117 (1995)
[erratum: Phys. Rev. Lett. \textbf{75}, 4155 (1995)], 
doi:10.1103/PhysRevLett.74.3113
%30 citations counted in INSPIRE as of 22 Jan 2021



%\cite{Aaij:2019kss}
\bibitem{Aaij:2019kss}
R.~Aaij \textit{et al.} [LHCb],
%``Observation of the doubly Cabibbo-suppressed decay $\Xi_{c}^{+}\to p\phi$,''
JHEP \textbf{04}, 084 (2019)
doi:10.1007/JHEP04(2019)084
[arXiv:1901.06222 [hep-ex]].
%4 citations counted in INSPIRE as of 23 Mar 2021

%\cite{Aaij:2019lwg}
\bibitem{Aaij:2019lwg}
R.~Aaij \textit{et al.} [LHCb],
%``Precision measurement of the $\Lambda_c^+$, $\Xi_c^+$ and $\Xi_c^0$ baryon lifetimes,''
Phys. Rev. D \textbf{100}, no.3, 032001 (2019)
doi:10.1103/PhysRevD.100.032001
[arXiv:1906.08350 [hep-ex]].
%19 citations counted in INSPIRE as of 23 Mar 2021
%\cite{Aaij:2020wtg}
\bibitem{Aaij:2020wtg}
R.~Aaij \textit{et al.} [LHCb],
%``First branching fraction measurement of the suppressed decay $\Xi_c^0\to \pi^-\Lambda_c^+$,''
Phys. Rev. D \textbf{102}, no.7, 071101 (2020)
doi:10.1103/PhysRevD.102.071101
[arXiv:2007.12096 [hep-ex]].
%1 citations counted in INSPIRE as of 23 Mar 2021

%\cite{Aaij:2020wil}
\bibitem{Aaij:2020wil}
R.~Aaij \textit{et al.} [LHCb],
%``Search for $CP$ violation in ${{{\varXi }} ^+_{c}} \rightarrow {p} {{K} ^-} {{\pi } ^+} $ decays using model-independent techniques,''
Eur. Phys. J. C \textbf{80}, no.10, 986 (2020)
doi:10.1140/epjc/s10052-020-8365-0
[arXiv:2006.03145 [hep-ex]].
%6 citations counted in INSPIRE as of 23 Mar 2021



%\cite{Li:2018qak}
\bibitem{Li:2018qak}
Y.~B.~Li \textit{et al.} [Belle],
%``First Measurements of Absolute Branching Fractions of the $\Xi_c^0$ Baryon at Belle,''
Phys. Rev. Lett. \textbf{122},   082001 (2019), 
doi:10.1103/PhysRevLett.122.082001
[arXiv:1811.09738 [hep-ex]].
%29 citations counted in INSPIRE as of 22 Jan 2021

\bibitem{ALICE}
J. Zhu on behalf of the ALICE collaboration,  PoS  {\bf ICHEP2020} (2021) 524.

%\cite{2103.06496}
\bibitem{2103.06496}
Y.~B.~Li \textit{et al.} [Belle],
%``Measurements of the branching fractions of semileptonic decays $\Xi_{c}^{0} \to \Xi^{-} \ell^{+} \nu_{\ell}$ and asymmetry parameter of $\Xi_{c}^{0} \to \Xi^{-} \pi^{+}$ decay,''
[arXiv:2103.06496 [hep-ex]].
%0 citations counted in INSPIRE as of 12 Mar 2021




%\cite{Zhao:2018zcb}
\bibitem{Zhao:2018zcb}
Z.~X.~Zhao,
%``Weak decays of heavy baryons in the light-front approach,''
Chin. Phys. C \textbf{42},  093101 (2018), 
doi:10.1088/1674-1137/42/9/093101
[arXiv:1803.02292 [hep-ph]].
%32 citations counted in INSPIRE as of 03 Feb 2021

%\cite{Liu:2010bh}
\bibitem{Liu:2010bh}
Y.~L.~Liu and M.~Q.~Huang,
%``A light-cone QCD sum rule approach for the $\Xi$ baryon electromagnetic form factors and the semileptonic decay Xi\_c ---\ensuremath{>} Xi e\textasciicircum{}+ nu\_e,''
J. Phys. G \textbf{37}, 115010 (2010), 
doi:10.1088/0954-3899/37/11/115010
[arXiv:1102.4245 [hep-ph]].
%8 citations counted in INSPIRE as of 22 Jan 2021

%\cite{Azizi:2011mw}
\bibitem{Azizi:2011mw}
K.~Azizi, Y.~Sarac and H.~Sundu,
%``Light cone QCD sum rules study of the semileptonic heavy $\Xi_{Q}$ and $\Xi'_{Q}$ transitions to $\Xi$ and $\Sigma $ baryons,''
Eur. Phys. J. A \textbf{48}, 2 (2012), 
doi:10.1140/epja/i2012-12002-1
[arXiv:1107.5925 [hep-ph]].
%18 citations counted in INSPIRE as of 26 Feb 2021

%\cite{Geng:2018plk}
\bibitem{Geng:2018plk}
C.~Q.~Geng, Y.~K.~Hsiao, C.~W.~Liu and T.~H.~Tsai,
%``Antitriplet charmed baryon decays with SU(3) flavor symmetry,''
Phys. Rev. D \textbf{97}, no.7, 073006 (2018), 
doi:10.1103/PhysRevD.97.073006
[arXiv:1801.03276 [hep-ph]].
%39 citations counted in INSPIRE as of 26 Feb 2021

%\cite{Geng:2020gjh}
\bibitem{Geng:2020gjh}
C.~Q.~Geng, C.~W.~Liu and T.~H.~Tsai,
%``Semileptonic weak decays of anti-triplet charmed baryons in the light-front formalism,''
[arXiv:2012.04147 [hep-ph]].
%0 citations counted in INSPIRE as of 11 Feb 2021

%\cite{Zhao:2021sje}
\bibitem{Zhao:2021sje}
Z.~X.~Zhao,
%``Semi-leptonic form factors of $\Xi_{c}\to\Xi$ in QCD sum rules,''
[arXiv:2103.09436 [hep-ph]].
%0 citations counted in INSPIRE as of 23 Mar 2021
%\cite{Borsanyi:2012zs}
\bibitem{Borsanyi:2012zs}
S.~Borsanyi, S.~D\"urr, Z.~Fodor, C.~Hoelbling, S.~D.~Katz, S.~Krieg, T.~Kurth, L.~Lellouch, T.~Lippert and C.~McNeile, \textit{et al.}
%``High-precision scale setting in lattice QCD,''
JHEP \textbf{09}, 010 (2012), 
doi:10.1007/JHEP09(2012)010
[arXiv:1203.4469 [hep-lat]].
%311 citations counted in INSPIRE as of 22 Feb 2021

%\cite{Zyla:2020zbs}
\bibitem{Zyla:2020zbs}
P.~A.~Zyla \textit{et al.} [Particle Data Group],
%``Review of Particle Physics,''
PTEP \textbf{2020},  083C01 (2020), 
doi:10.1093/ptep/ptaa104
%830 citations counted in INSPIRE as of 25 Feb 2021


\bibitem{supplemental}
 Supplementary material of this work.
 
%\cite{Chang:2018uxx}
\bibitem{Chang:2018uxx}
C.~C.~Chang, A.~N.~Nicholson, E.~Rinaldi, E.~Berkowitz, N.~Garron, D.~A.~Brantley, H.~Monge-Camacho, C.~J.~Monahan, C.~Bouchard and M.~A.~Clark, \textit{et al.}
%``A per-cent-level determination of the nucleon axial coupling from quantum chromodynamics,''
Nature \textbf{558},   91-94 (2018), 
doi:10.1038/s41586-018-0161-8
[arXiv:1805.12130 [hep-lat]].
%99 citations counted in INSPIRE as of 16 Feb 2021


%\cite{Bourrely:2008za}
\bibitem{Bourrely:2008za}
C.~Bourrely, I.~Caprini and L.~Lellouch,
%``Model-independent description of B ---\ensuremath{>} pi l nu decays and a determination of |V(ub)|,''
Phys. Rev. D \textbf{79}, 013008 (2009)
[erratum: Phys. Rev. D \textbf{82}, 099902 (2010)], 
doi:10.1103/PhysRevD.82.099902
[arXiv:0807.2722 [hep-ph]].
%298 citations counted in INSPIRE as of 19 Feb 2021

%\cite{Edwards:2004sx}
\bibitem{Edwards:2004sx}
R.~G.~Edwards \textit{et al.} [SciDAC, LHPC and UKQCD],
%``The Chroma software system for lattice QCD,''
Nucl. Phys. B Proc. Suppl. \textbf{140}, 832 (2005), 
doi:10.1016/j.nuclphysbps.2004.11.254
[arXiv:hep-lat/0409003 [hep-lat]].
%761 citations counted in INSPIRE as of 16 Feb 2021

%\cite{Clark:2009wm}
\bibitem{Clark:2009wm}
M.~A.~Clark, R.~Babich, K.~Barros, R.~C.~Brower and C.~Rebbi,
%``Solving Lattice QCD systems of equations using mixed precision solvers on GPUs,''
Comput. Phys. Commun. \textbf{181}, 1517-1528 (2010), 
doi:10.1016/j.cpc.2010.05.002
[arXiv:0911.3191 [hep-lat]].
%246 citations counted in INSPIRE as of 12 Feb 2021

%\cite{Babich:2011np}
\bibitem{Babich:2011np}
R.~Babich, M.~A.~Clark, B.~Joo, G.~Shi, R.~C.~Brower and S.~Gottlieb,
%``Scaling Lattice QCD beyond 100 GPUs,''
doi:10.1145/2063384.2063478
[arXiv:1109.2935 [hep-lat]].
%85 citations counted in INSPIRE as of 09 Feb 2021

%\cite{Clark:2016rdz}
\bibitem{Clark:2016rdz}
M.~A.~Clark, B.~Jo\'o, A.~Strelchenko, M.~Cheng, A.~Gambhir and R.~Brower,
%``Accelerating Lattice QCD Multigrid on GPUs Using Fine-Grained Parallelization,''
[arXiv:1612.07873 [hep-lat]].
%23 citations counted in INSPIRE as of 09 Feb 2021

%\cite{Bi:2020wpt}
\bibitem{Bi:2020wpt}
Y.~J.~Bi, Y.~Xiao, W.~Y.~Guo, M.~Gong, P.~Sun, S.~Xu and Y.~B.~Yang,
%``Lattice QCD package GWU-code and QUDA with HIP,''
PoS \textbf{LATTICE2019}, 286 (2020), 
doi:10.22323/1.363.0286
[arXiv:2001.05706 [hep-lat]].
%5 citations counted in INSPIRE as of 11 Feb 2021




\end{thebibliography}
\end{document}